\documentclass[aps,groupedaddress,floatfix]{revtex4}
\usepackage{amssymb}
\usepackage{amsmath}
\usepackage{graphicx}
\usepackage{color}
\DeclareMathOperator{\sgn}{sgn}

\begin{document}

\vspace{5mm}

\title{Zero temperature limit for (1+1) directed polymers
       with correlated random potential}

\author{Victor Dotsenko}

\affiliation{Universit\'e Paris VI, 75252 Paris, France}

\affiliation{ Landau Institute for Theoretical Physics, Moscow, Russia}

\date{\today}

\begin{abstract}
Zero temperature limit in (1+1) directed polymers with finite range correlated random potential is studied.
In terms of the standard replica technique it is demonstrated that in this limit  the considered system 
reveals  the one-step replica symmetry breaking structure  similar to the one which 
takes place in the  Random Energy Model. In particular, it is shown that at the temperature
$T_{*} \sim (u R)^{1/3}$ (where $u$ and $R$ are the strength and the correlation length of the random
potential) there is a {\it crossover} from the high- to the low-temperature regime.
Namely, in the high-temperature regime at $T \gg T_{*}$ the model is equivalent to the one with the 
$\delta$-correlated potential (in which the non-universal prefactor
of the free energy is proportional to $T^{-2/3}$), 
while at $T \ll T_{*}$ the non-universal prefactor of the free energy saturates 
at a finite (temperature independent) value.

\end{abstract}

\maketitle

\medskip

\hfill  {\bf \large \it In the memory of Sergei Korshunov}

\section{Introduction}

In this paper we study the statistical properties of one-dimensional directed polymers
in correlated random potential in the zero-temperature limit. This system is defined
in terms of the so called $(1+1)$ Hamiltonian
\begin{equation}
   \label{1}
   H[\phi(\tau), V] = \int_{0}^{t} d\tau
   \Bigl\{\frac{1}{2} \bigl[\partial_\tau \phi(\tau)\bigr]^2
   + V[\phi(\tau),\tau]\Bigr\};
\end{equation}
where $\phi(\tau)$ is a scalar field defined
within an interval $0 \leq \tau \leq t$ and
$V(\phi,\tau)$ is the Gaussian distributed random potential with a zero mean, $\overline{V(\phi,\tau)}=0$,
and the correlation function
\begin{equation}
\label{2}
\overline{V(\phi,\tau)V(\phi',\tau')} = u \delta(\tau-\tau') U(\phi-\phi')
\end{equation}
Here the parameter $u$ is the strength of the disorder and $U(\phi)$ is the "spatial" correlation
function characterized by the correlation length $R$. For simplicity we take
\begin{equation}
\label{3}
U(\phi) \; = \; \frac{1}{\sqrt{2\pi} \, R} \; \exp\Bigl\{-\frac{\phi^{2}}{2 R^{2}}\Bigr\}
\end{equation}

This problem, is equivalent to the one of the KPZ-equation
\cite{KPZ} describing the time evolution of an interface
in the presence of noise. The case of the $\delta$-correlated random potential
(it is recovered in the limit $R \to 0$ in eq.(\ref{3}))
has been the focus  of intense studies during past three
decades
\cite{hhf_85,numer1,numer2,kardar_87,bouchaud-orland,hh_zhang_95,
Johansson,Prahofer-Spohn,Ferrari-Spohn1,KPZ-TW1a,KPZ-TW1b,KPZ-TW1c,KPZ-TW2,BA-TW2,BA-TW3,
LeDoussal1,LeDoussal2,goe,LeDoussal3,Corwin,Borodin}.
At present it is well established that in this case the fluctuations of the free energy
of the model, eqs.(\ref{1})-(\ref{3}), (with $R \to 0$) are described by the Tracy-Widom  (TW)
distribution \cite{TW-GUE}. It can be argued (see Section II) that the model, eqs.(\ref{1})-(\ref{3}),
with {\it finite} correlation length $R$ in the high temperature limit is equivalent to the one
with the $\delta$-correlated random potentials. Note that this provides the basis for the
short-length regularization of the model with the $\delta$-potentials which, in fact,
is ill defined at short scales.

The zero-temperature limit for this model is much less clear. The problem is that in the exact solution of the 
model with $\delta$-correlated potentials (see e.g. \cite{KPZ-TW2,BA-TW2,BA-TW3,LeDoussal1}) the fluctuating part 
of the free energy $F(T, t)$ (in the limit $t\to\infty$) is proportional to $(u/T)^{2/3} \, t^{1/3} \, f$ 
(where $T$ is the temperature and the random quantity 
$f \sim 1$ is described by the TW distribution) and this free energy does not reveal any finite zero-temperature limit.
The physical origin of this pathology is clear: the exact solution mentioned above is valid {\it only}
for the model with $\delta$-correlated potentials which is ill defined at short scales while it is the 
short scales which are getting the most relevant in the limit $T \to 0$. One can propose two types of the 
regularization of the model on short scales: (1) introducing a lattice, and (2) keeping continuous structure of
the "space-time" but introducing smooth finite size correlations for the random potential like in eq.(\ref{3}). 
In both cases, however, the solution derived for the model with $\delta$-correlated potentials becomes
not valid. Nevertheless, it is generally believed that in the zero-temperature  
limit the considered system must reveal the same TW distribution. On one hand there are exact results
for strictly zero temperature lattice models revealing main features of (1+1) directed polymers (see e.g
rigorous solution of the directed polymer lattice model with geometric disorder \cite{Johansson}).
On the other hand there are analytic indications that in the zero-temperature limit 
the continuous model with finite correlation length of the type introduced above, eqs.(\ref{1})-(\ref{3}), 
after crossover to a new regime keeps the main features of the finite-temperature solution \cite{korshunov,Lecomte}.

This paper (following an old idea first formulated in the paper \cite{korshunov}) is an attempt to formulate a general 
scheme which would allow to obtain a finite  zero-temperature limit for the continuous model  (\ref{1})-(\ref{3}). 
In terms of the standard replica technique it is demonstrated that in the limit $T \to 0$ the considered system 
reveals  the one-step replica symmetry breaking structure  similar to the one which 
takes place in the low-temperature phase of the Random Energy Model (REM) \cite{rem}. 
Of course, the considered system (unlike REM) reveal no phase transition: here at the temperature
$T_{*} \sim (u R)^{1/3}$ we observe only a {\it crossover}
from the high- to the low-temperature regime. Namely, at $T \gg T_{*}$ the fluctuating part of the free energy is
the same as in the model with the $\delta$-interactions, $F(T, t) \simeq (u/T)^{2/3} \, t^{1/3} \, f$,
while at $T \ll T_{*}$ the non-universal prefactor of the free energy saturates at finite
value, $F(T=0, t) \simeq (u^{2}/R)^{1/9} \, t^{1/3} \, f$. The probability distribution function of the
random quantity $f$ is, of course, expected to be the TW one, although at present stage this is not proved yet.

\section{Replica formalism}

For a given realization of the random potential $ V[\phi,\tau]$
the partition function of the considered system (with fixed boundary conditions) is
\begin{equation}
\label{4}
Z(t) = \int_{\phi(0)=0}^{\phi(t)=0} {\cal D}\phi(\tau)
        \exp\bigl\{-\beta H[\phi(\tau), V]\bigr\} \; = \; \exp\bigl\{-\beta F(t)\bigr\}
\end{equation}
where $\beta$ is the inverse temperature,
$F(t)$ is the free energy which is a random quantity and the integration is taken over all
trajectories $\phi(\tau)$ with the zero boundary conditions at $\tau = 0$ and $\tau = t$.
The free energy probability distribution function
$P(F)$ can be studied in terms of the integer moments of the above partition function:
\begin{equation}
\label{5}
\overline{Z^{N}} \equiv Z(N,t) \; = \; \prod_{a=1}^{N}\int_{\phi_{a}(0)=0}^{\phi_{a}(t)=0}
        {\cal D}\phi_{a}(\tau) \;
        \overline{
        \Biggl(
        \exp\Bigl\{-\beta \sum_{a=1}^{N} H[\phi_{a}(\tau), V]\Bigr\}
        \Biggr)}
        \; = \;
        \int_{-\infty}^{+\infty} dF \, P(F) \, \exp\bigl\{-\beta N F\bigr\}
\end{equation}
where $\overline{(...)}$ denotes the averaging over the random potentials $V[\phi,\tau]$.
The quantity $Z(N, t)$ is called the replica partition function. After performing simple Gaussian
averaging and using eq.(\ref{2}) we get
\begin{equation}
\label{6}
Z(N,t) \; = \; \prod_{a=1}^{N}\int_{\phi_{a}(0)=0}^{\phi_{a}(t)=0}
        {\cal D}\phi_{a}(\tau) \;
        \exp\Bigl\{-\beta  H_{N}[\boldsymbol{\phi}(\tau)]\Bigr\}
\end{equation}
where
$\boldsymbol{\phi}(\tau) \equiv \bigl(\phi_{1}(\tau), \, \phi_{2}(\tau), \, ... \, , \phi_{N}(\tau)\bigr)$
and
\begin{equation}
   \label{7}
   \beta H_{N}[\boldsymbol{\phi}] =  \int_{0}^{t} d\tau
   \Biggl[
   \frac{1}{2} \beta \sum_{a=1}^{N} \Bigl(\partial_\tau \phi_{a}(\tau)\Bigr)^2
   \; - \;
   \frac{1}{2} \beta^{2} \, u \,
    \sum_{a,b=1}^{N} U\bigl(\phi_{a}(\tau) - \phi_{b}(\tau)\bigr)
   \Biggr];
\end{equation}
is the replica Hamiltonian which describes $N$ elastic strings
$\bigl\{\phi_{1}(\tau), \, \phi_{2}(\tau), \, ... \, , \phi_{N}(\tau)\bigr\}$
with the finite width attractive interactions $U\bigl(\phi_{a} - \phi_{b}\bigr)$,
eq.(\ref{3}).

To compute the replica partition function $Z(N,t)$, eq.(\ref{6}), one introduces
more complicated object:
\begin{equation}
\label{8}
\Psi(x_{1}, \, x_{2}, \, ... \, x_{N}; \; t) \; = \;
        \prod_{a=1}^{N}\int_{\phi_{a}(0)=0}^{\phi_{a}(t)=x_{a}}
        {\cal D}\phi_{a}(\tau) \;
        \exp\Bigl\{-\beta  H_{N}[\boldsymbol{\phi}(\tau)]\Bigr\}
\end{equation}
such that $Z(N,t) \, = \, \Psi(x_{1},  x_{2},  ...  x_{N}; \, t)\big|_{x_{a}=0}$.
One can easily show that $\Psi({\bf x}; \, t)$ is the wave function of N quantum bosons defined by
the imaginary time Schr\"odinger equation
\begin{equation}
\label{9}
\beta \frac{\partial}{\partial t} \Psi({\bf x}; \, t) \; = \;
\frac{1}{2}\sum_{a=1}^{N} \, \frac{\partial^{2}}{\partial x_{a}^{2}} \Psi({\bf x}; \, t)
\; + \; \frac{1}{2} \, \beta^{3} u \, \sum_{a,b=1}^{N} U(x_{a} - x_{b}) \, \Psi({\bf x}; \, t)
\end{equation}
with the initial condition
\begin{equation}
\label{10}
\Psi({\bf x}; \, 0) \; = \; \prod_{a=1}^{N}\, \delta(x_{a})
\end{equation}
In the high temperature limit ($\beta \to 0$) the typical distance between the particles (defined
by the wave function $\Psi({\bf x}; \, t)$) is much larger that the size $R$ of the potential
$U(x)$, eq.(\ref{3}) (see below). In this case the potential $U(x)$ can be approximated by
the $\delta$-function, so that a generic solution of the Schr\"odinger equation (\ref{9})
(with $U(x) = \delta(x)$) is obtained in terms of the Bethe ansatz eigenfunctions \cite{Lieb-Liniger,McGuire,Yang}. A generic eigenstate of the system described by eq.(\ref{9})
(with $U(x) = \delta(x)$) is  defined by $N$ momenta $\{Q_{a}\} \; \; (a = 1, ... N)$
which splits into $M$ "clusters" ($1 \leq M \leq N$) each defined by continuous real momenta
$q_{\alpha} \; \; (\alpha = 1, ... , M)$ and $n_{\alpha}$ discrete imaginary components
(for details see e.g. \cite{BA-TW3,rev-TW}):
\begin{equation}
\label{11}
Q_{a} \; \to \; q_{\alpha}^{r} \, = \,
                q_{\alpha} \, - \,
                \frac{i}{2}  \kappa \bigl(n_{\alpha} + 1 - 2 r\bigr) \, ;
                \; \; \;
                r = 1, ... , n_{\alpha} \, ;
                \; \; \;
                \alpha = 1, ... , M
\end{equation}
(where $\kappa = \beta^{3} u$) with the global constraint $\sum_{\alpha=1}^{M} n_{\alpha} \, = \, N$.
Explicitly, the corresponding eigenfunction $\Psi_{Q}({\bf x})$ reads
\begin{equation}
\label{12}
\Psi_{{\bf Q}}({\bf x}) \; = \; \sum_{{\cal P}\in S_{N}}
                          \prod_{a < b}^{N}
                          \Biggl[1 + i\kappa \, \frac{\sgn(x_{a}-x_{b})}{
                                                 Q_{{\cal P}_{a}} - Q_{{\cal P}_{b}}}
                          \Biggr] \,
                          \exp\Biggl\{i\sum_{a=1}^{N} Q_{{\cal P}_{a}} \, x_{a}\Biggr\}
\end{equation}
where the vector ${\bf Q}$ denotes the set of all $N$ momenta eq.(\ref{11}) and
the summation goes over $N!$ permutations ${\cal P}$ of $N$ momenta $Q_{a}$,
over $N$ particles $x_{a}$. In particular, for the ground state wave function
($M = 1; \; Q_{a} = q - \frac{i}{2} \kappa (N + 1 - 2a)$)  one gets:
\begin{equation}
\label{13}
\Psi_{q}({\bf x}) \; \propto \;
   \exp\Bigl\{-\frac{1}{4} \, \kappa \sum_{a,b=1}^{N} |x_{a} - x_{b}| \, + \, i q \sum_{a=1}^{N} x_{a}
   \Bigr\}
\end{equation}
For the excited states (with $M > 1$) the generic wave function can be represented as a linear combination of various products of the "cluster" wave functions which have the structure similar to the one
in eq.(\ref{13}) \cite{BA-TW3}. We see that in any case the typical distance between the particles
is of the order of $\kappa^{-1} \, = \, (\beta^{3} u)^{-1}$. Thus, the approximation of the potential
$U(x)$, eq.(\ref{3}), by the $\delta$-function is justified provided $(\beta^{3} u)^{-1} \gg R$
which for a given $R$ and $u$ is valid only in the high temperature region,
\begin{equation}
\label{14}
\beta \, \ll \, (R u)^{-1/3}
\end{equation}
At temperatures of the order of the $(R u)^{1/3}$ and below the typical distance between particles becomes
comparable with the size $R$ of the potential $U(x)$, and therefore its approximation by the
$\delta$-function is no longer valid, which makes the considered model unsolvable (at least for
 the time being). It turns out, however, that in the zero temperature limit,
at $T \ll (R u)^{1/3}$ the situation somewhat simplifies again (see next section).

\section{Zero temperature limit}

One can easily show that the parameters of the considered system can be redefined in such a way
that  in the
limit $t \to \infty$  the properties of the system would depend on the
only parameter, which is the reduced temperature $\tilde{T} \, = \, T/T_{*}$ where
\begin{equation}
\label{15}
T_{*} \, = \, \Bigl(\frac{uR}{\sqrt{2\pi}}\Bigr)^{1/3}
\end{equation}
Indeed, redefining
\begin{eqnarray}
\nonumber
\phi &\to & R \, \phi
\\
\beta & = & T_{*}^{-1} \, \tilde{\beta}
\label{16}
\\
\nonumber
\tau & \to & \tau_{*} \, \tau
\end{eqnarray}
with
\begin{equation}
\label{17}
\tau_{*} \, = \, \Bigl(\sqrt{2\pi} \, R^{5} u^{-1} \Bigr)^{1/3}
\end{equation}
instead of the replica Hamiltonian, eq.(\ref{7}), one gets
\begin{equation}
   \label{18}
   \beta \tilde{H}_{N}[\boldsymbol{\phi}] =  \int_{0}^{t/\tau_{*}} d \tau
   \Biggl[
   \frac{1}{2} \, \tilde{\beta} \sum_{a=1}^{N} \Bigl(\partial_\tau \phi_{a}(\tau)\Bigr)^2
   \; - \;
   \frac{1}{2} \, \tilde{\beta}^{2}
    \sum_{a,b=1}^{N} U_{0}\bigl(\phi_{a}(\tau) - \phi_{b}(\tau)\bigr)
   \Biggr];
\end{equation}
where
\begin{equation}
\label{19}
U_{0}(\phi) \; = \;  \exp\Bigl\{-\frac{1}{2} \, \phi^{2} \Bigr\}
\end{equation}
Accordingly, the  wave function $\tilde{\Psi}({\bf x}; t)$ defined by eq.(\ref{8}) with
$\beta H_{N}[\boldsymbol{\phi}]$ replaced by $\beta \tilde{H}_{N}[\boldsymbol{\phi}]$
is given by the solution of the Schr\"odinger equation
\begin{equation}
\label{20}
2 \tilde{\beta} \frac{\partial}{\partial \tilde{t}} \tilde{\Psi}({\bf x}; \, \tilde{t}) \; = \;
\sum_{a=1}^{N} \, \frac{\partial^{2}}{\partial x_{a}^{2}} \tilde{\Psi}({\bf x}; \, \tilde{t})
\; + \;  {\tilde{\beta}}^{3} \sum_{a,b=1}^{N} U_{0}(x_{a} - x_{b}) \,
\tilde{\Psi}({\bf x}; \, \tilde{t})
\end{equation}
where $\tilde{t} = t/\tau_{*}$. The corresponding eigenvalue equation for the eigenfunctions
$\psi({\bf x})$, defined by the relation
$\tilde{\Psi}({\bf x}; \, \tilde{t}) \, = \, \psi({\bf x}) \exp\{- \tilde{t} E\}$,
reads:
\begin{equation}
\label{21}
- 2 E \, \tilde{\beta} \, \psi({\bf x}) \; = \;
\sum_{a=1}^{N} \, \frac{\partial^{2}}{\partial x_{a}^{2}} \psi({\bf x})
\; + \;  {\tilde{\beta}}^{3} \sum_{a,b=1}^{N} U_{0}(x_{a} - x_{b}) \, \psi({\bf x})
\end{equation}
As in what follows we will be interested only in the limit
$t \to \infty$ the rescaling  of the time $t$  by the factor $\tau_{*}$ does not change anything.
In other words, in the limit $t \to \infty$ our problem
(defined by the replica Hamiltonian (\ref{18}) or the Schr\"odinger equation (\ref{20}))
is controlled by the only parameter $\tilde{\beta}$.

\vspace{5mm}

In the  zero temperature limit ($\tilde{\beta} \to \infty$),
according to the discussion at the end of the previous section,
one may naively suggest that the typical distance between particles defined by the
eigenfunctions $\psi({\bf x})$, eq.(\ref{21}), becomes small compared to size ($\sim 1$)
of the potential $U_{0}(x)$. In other words, all the particles are expected to be localized
near the bottom of this potential so that one could approximate
\begin{equation}
\label{22}
U_{0}(\phi) \; \simeq \;  1 \; - \; \frac{1}{2} \, x^{2}
\end{equation}
For the  corresponding eigenvalue equation
\begin{equation}
\label{23}
- 2 E \, \tilde{\beta} \, \psi({\bf x}) \; = \;
\sum_{a=1}^{N} \, \frac{\partial^{2}}{\partial x_{a}^{2}} \psi({\bf x})
\; + \;  {\tilde{\beta}}^{3} N^{2} \psi({\bf x}) \; - \;
\frac{1}{2} \, {\tilde{\beta}}^{3} \sum_{a,b=1}^{N} (x_{a} - x_{b})^{2} \, \psi({\bf x})
\end{equation}
one finds simple (exact) ground state solution
\begin{equation}
\label{24}
\psi_{0}({\bf x}) \; = \;
C \, \exp\Biggl\{
-\frac{1}{4} \, \sqrt{\frac{{\tilde{\beta}}^{3}}{N}} \sum_{a,b=1}^{N} (x_{a} - x_{b})^{2}
\Biggr\}
\end{equation}
where $C$ is the normalization constant, and
\begin{equation}
\label{25}
E_{0}(\tilde{\beta}, N) \; = \; -\frac{1}{2} \, \bigl(\tilde{\beta} \, N\bigr)^{2}
              +\frac{1}{2} \, (N-1) \, \sqrt{\tilde{\beta} \, N}
\end{equation}
Using the explicit form of the wave function (\ref{24}) one can easily estimate the average distance
$\Delta x$ between arbitrary two particles in this $N$-particle system:
\begin{equation}
\label{26}
\Delta x \; \sim \; \sqrt{\frac{1}{N(N-1)} \sum_{a,b=1}^{N}\overline{(x_{a} - x_{b})^{2}}}
\; \sim \;
\bigl(\tilde{\beta}^{3} N\bigr)^{-1/4}
\end{equation}
In the limit $\tilde{\beta} \to \infty$ (at fixed $N$)
$\Delta x \sim \tilde{\beta}^{-3/4} \to 0$. Therefore, in the zero temperature limit
for any fixed value of $N$ the wave function
$\psi_{0}({\bf x})$ is indeed nicely localized neat the "bottom" of the potential well
$U_{0}(x)$ which justifies the approximation (\ref{22}).
On the other hand, one can easily see that both the wave function (\ref{24}) and the
ground state energy (\ref{25}) demonstrate completely
pathological behavior in the limit $N\to 0$ which is crucial for the reconstruction of the
free energy distribution function in the limit $t \to \infty$.
The typical size $\Delta x$ (\ref{26}) of the wave function (\ref{24}) grows with decreasing
$N$ and become of the order of one at $N \sim  \tilde{\beta}^{-3}$. Therefore at smaller values
of $N$ the ground state wave function must have essentially different form.
The simplest way to let the particles enjoy their mutual attraction
while keeping their number in the well finite consists in splitting them into several well separated
groups, such that each group would consist of finite number of particles
This idea (similar to the one step replica symmetry breaking solution of the Random Energy Model
\cite{rem}) has been proposed by Sergei Korshunov many years ago \cite{korshunov}.
More specifically,
let us split $N$ particles into $K$ groups each consisting of $m$ particles, so that $K = N/m$.
In this way, instead of the coordinates of the particles $\{x_{a}\} \; \; (a=1,...,N)$
we introduce the coordinates of the center of masses of the groups
$\{X_{\alpha}\} \; \; (\alpha = 1, ..., K)$ and the deviations
$\xi_{i}^{\alpha} \; \; (i = 1, ... , m)$ of the particles of a given  group $\alpha$
from the position of its center of mass:
\begin{equation}
\label{27}
x_{a} \; \to \; X_{\alpha} \; + \; \xi_{\alpha}^{i} \, , \; \; \; \; \alpha = 1, ..., K
                                                    \, , \; \; \; \; i = 1, ..., m
                                                    \, , \; \; \; \; K = N/m
\end{equation}
where $\sum_{i=1}^{m} \xi_{\alpha}^{i} \, = \, 0$. It is supposed that in the zero temperature limit
 ($\tilde{\beta} \to \infty$)  the typical value
of the particle's deviations inside the groups are small,
$\langle \bigl(\xi_{i}^{\alpha}\bigr)^{2}\rangle\big|_{\tilde{\beta} \to \infty} \, \to \, 0$,
while the value of the typical distance between the groups remains finite.

In terms of the above ansatz the original replica partition function
\begin{equation}
\label{28}
Z(N,\tilde{t}) \; = \; \prod_{a=1}^{N}\int_{\phi_{a}(0)=0}^{\phi_{a}(\tilde{t})=0}
        {\cal D}\phi_{a}(\tau) \;
        \exp\Biggl\{
        -\frac{1}{2}\int_{0}^{\tilde{t}} d \tau
   \Biggl[
   \tilde{\beta} \sum_{a=1}^{N} \Bigl(\partial_\tau \phi_{a}\Bigr)^2
   \; - \;
   \tilde{\beta}^{2}
    \sum_{a,b=1}^{N} U_{0}\bigl(\phi_{a} - \phi_{b}\bigr)
   \Biggr]
        \Biggr\}
\end{equation}
factorizes into two independent contributions: (a) the "internal" partition functions
$Z_{0}\bigl(\tilde{\beta}, m, \tilde{t}\bigr)$  of tightly bound groups
of $m$ polymers for which one can use the approximation (\ref{22}), and (b)
the "external" partition function ${\cal Z}\bigl(\tilde{\beta}, N/m, m, \tilde{t}\bigr)$
of $N/m$ "complex" polymers (each consisting of $m$ tightly bound original polymers):
\begin{equation}
\label{29}
Z(N,\tilde{t}) \; \simeq \;
\Bigl[Z_{0}\bigl(\tilde{\beta}, m, \tilde{t}\bigr)\Bigr]^{\frac{N}{m}} \times
{\cal Z}\bigl(\tilde{\beta}, N/m, m, \tilde{t}\bigr)
\end{equation}
In the limit $\tilde{t} \to \infty$ the "internal" partition functions
$Z_{0}\bigl(\tilde{\beta}, m, \tilde{t}\bigr)$ is dominated by the ground state,
eqs.(\ref{24})-(\ref{25}):
\begin{equation}
\label{30}
Z_{0}\bigl(\tilde{\beta}, m, \tilde{t}\bigr) \; \simeq \;
\exp\bigl\{- E_{0}(\tilde{\beta}, m) \, \tilde{t} \bigr\} \times
\psi_{0}(\boldsymbol{\xi})\big|_{\boldsymbol{\xi}=0}
\; \propto \;
\exp\Biggl\{
\frac{1}{2} \,\Bigl[ \bigl(\tilde{\beta} m\bigr)^{2}
              +(1-m) \, \sqrt{\tilde{\beta} m} \; \Bigr] \, \tilde{t}
\Biggr\}
\end{equation}
According to the definition (\ref{28}), the "external" partition function
${\cal Z}\bigl(\tilde{\beta}, N/m, m, \tilde{t}\bigr)$
can be represented as follows:
\begin{equation}
\label{31}
{\cal Z}\bigl(\tilde{\beta}, N/m, m, \tilde{t}\bigr) \; = \; \prod_{\alpha=1}^{N/m}\int_{\phi_{\alpha}(0)=0}^{\phi_{\alpha}(\tilde{t})=0}
        {\cal D}\phi_{\alpha}(\tau) \;
        \exp\Biggl\{
        -\frac{1}{2}\int_{0}^{\tilde{t}} d \tau
   \Biggl[
   (\tilde{\beta}m) \, \sum_{\alpha=1}^{N/m} \Bigl(\partial_\tau \phi_{\alpha}\Bigr)^2
   \; - \;
   (\tilde{\beta} m)^{2}
    \sum_{\alpha\not=\beta}^{N/m} U_{0}\bigl(\phi_{\alpha} - \phi_{\beta}\bigr)
   \Biggr]
        \Biggr\}
\end{equation}

Now, to make the present approach self-consistent one has to formulate the procedure
which would define the value of the parameter $m$ which for the moment remains arbitrary.
In fact, the algorithm which would fix the value of this parameter can be borrowed 
from the standard procedure of replica symmetry breaking (RSB) scheme of spin glasses.
In particular, it is successfully used in the one-step RSB solution of the Random Energy Model (REM)
\cite{rem}. In simple words, this procedure is in the following. Originally the parameter $m$ is introduced 
as an  integer bounded by the condition $1 \leq m \leq N$ and  such that the replica parameter $N$
must be a multiple of $m$ as the number of the groups of particles $K = N/m$ must be an integer. 
After analytic continuation of the replica parameters $N$ and $m$ to arbitrary real values in the 
limit $N \to 0$ the above restriction $1 \leq m \leq N$ (interpreted now as $m$ is bounded 
{\it between} $1$ and $N$) turns into $N \leq m \leq 1$. Then the physical value of the parameter $m_{*}$
is fixed by the condition that the extensive part of the free energy of the considered
system $F_{0}(m)$ (considered as a function of $m$ at the interval $N \leq m \leq 1$) 
has the {\it maximum} at $m = m_{*}$. Of course, from the mathematical point of view this 
procedure is ill grounded (or better to say not grounded at all), but somehow in all know
cases (where the solution can also be obtained in some other methods) it works perfectly well. 
In any case, for the moment we don't have any other method anyway.

In the present case the situation is even worse as 
the solution of the  problem defined by the replica partition function (\ref{31})
(which also gives a contribution to the extensive part of the free energy)
is not known. Nevertheless,
it would be natural to suggest that its generic structure is similar to the one
with the  $\delta$-interactions (instead of  $U_{0}(\phi)$).
Namely, let us suppose (like in the problem with the  $\delta$-interactions) that
the free energy of the system which is defined by the replica partition function
(\ref{31}) contains both the linear in time (self-averaging) part $f_{0}(\tilde{\beta}m)$ and
the fluctuating part which scales with time as $t^{1/3}$. In this case, due to factorization,
eq.(\ref{29}), in the limit $\tilde{t} \to \infty$, the total free energy of the system $F$
can be represented as the sum of two contributions:
\begin{equation}
\label{32}
F \; = \; F_{0} \, t \; + \; {\cal F} \, t^{1/3}\
\end{equation}
where the fluctuating part ${\cal F}$ is defined by the solution of the problem (\ref{31}),
while the linear part $F_{0}$ is given by the sum of the "internal" contribution
$E_{0}(\tilde{\beta}, m)$, eq.(\ref{25}), and the contribution $f_{0}(\tilde{\beta}m)$ coming
from the "external" partition function (\ref{31}). In the limit $t \to \infty$ 
the fluctuating contribution of the free energy $\sim t^{1/3}$ can be neglected compared with 
its linear in $t$ part so that 
we gets
\begin{eqnarray}
 \nonumber
F_{0}(\tilde{\beta}, m) &=& - \lim_{t\to\infty} \, \frac{1}{\beta N t} \, \ln\Bigl[
Z_{0}^{N/m} \, {\cal Z} \Bigr]
\\
\nonumber
\\
\nonumber
&=&  
- \lim_{t\to\infty} \, \frac{T_{*}}{\tilde{\beta} N t} \, \Bigl[
\frac{N}{m} \, E_{0}(\tilde{\beta}, m) \, \frac{t}{\tau_{*}} \; + \; 
\frac{N}{m} \, f_{0}(\tilde{\beta}m) \frac{t}{\tau_{*}} 
\Bigr]
\\
\nonumber
\\
&=&
\frac{T_{*}}{2 \tau_{*}} \Bigl[
-(\tilde{\beta}m) \; - \; (1-m) \, (\tilde{\beta}m)^{-1/2} \; + \; 
\frac{2}{(\tilde{\beta}m)} f_{0}(\tilde{\beta}m) 
\Bigr]
 \label{33}
\end{eqnarray}
where the quantities $T_{*}$, $\tau_{*}$ and $E_{0}(\tilde{\beta}m)$ are defined in eqs.(\ref{15}), (\ref{17}) and (\ref{25}).
The parameter $m$ is defined by the solution of the equation
\begin{equation}
 \label{34}
\frac{\partial}{\partial m} \, F_{0}(\tilde{\beta}, m) \; = \; 0
\end{equation}
which corresponds to the {\it maximum} of the function $F_{0}(m)$ at the interval $0 \leq m \leq 1$
(as in the limit $t\to\infty$ the relevant values of the replica parameter $N$ which define
the statistics of the free energy fluctuations are of the order of $t^{-1/3} \to 0$).
Let us introduce  a new parameter 
\begin{equation}
 \label{35}
\zeta \; = \; \tilde{\beta} m
\end{equation}
which is supposed to remain {\it finite} in the zero-temperature limit,  $\tilde{\beta} \to \infty$.
In terms of this parameter the expression for the self-averaging free energy $F_{0}$, eq.(\ref{33}),
reduces to
\begin{equation}
 \label{36}
F_{0}(\zeta) \; \simeq \; \frac{T_{*}}{2 \tau_{*}} \Bigl[
-\zeta  \; - \;  \zeta^{-1/2} \; + \; 
\frac{2}{\zeta} \, f_{0}(\zeta) 
\Bigr]
\end{equation}
where  for a finite value of $\zeta$ the factor $(1-m)$ in second term in eq.(\ref{33}) 
in the limit $\tilde{\beta} \to \infty$ can be approximated
as $(1-m) \; = \; \bigl(1 \, - \, \zeta/\tilde{\beta}\bigr) \; \simeq \; 1$. 
Then, according to eq.(\ref{36}) in the zero-temperature  limit the  
saddle-point equation (\ref{34}) reads
\begin{equation}
 \label{37}
-1 \; + \; \frac{1}{2} \, \zeta^{-3/2} \; + \; \frac{d}{d \zeta} \, \Bigl[ \frac{2}{\zeta} f_{0}(\zeta) \Bigr] \; = \; 0
\end{equation}
which contains no parameters and correspondingly its solution defines a {\it finite} value of $\zeta_{*}$ (which is 
just a number). As a matter of illustration, if we approximate the solution of the "external problem", eq.(\ref{31}),
by the well known result for the model with the $\delta$-interactions, namely
$f_{0}(\tilde{\beta} m) \; = \; \frac{1}{24} \bigl(\tilde{\beta} m\bigr)^{5}$ (see e.g. \cite{BA-TW3}), 
the above equation  reduces to 
\begin{equation}
 \label{38}
-1 \; + \; \frac{1}{2} \, \zeta^{-3/2} \; + \; \frac{1}{3} \zeta^{3} \; = \; 0
\end{equation}
The solution of this equation is $\zeta_{*} \; \simeq \; 0.68$.

Thus, in the zero temperature limit the partition function of the "external problem", eq.(\ref{31}), as the function
of a new replica parameter $N/m \equiv K$ becomes temperature independent:
\begin{equation}
\label{39}
{\cal Z}\bigl(\zeta_{*}; K, \tilde{t}\bigr) \; = \; \prod_{\alpha=1}^{K}\int_{\phi_{\alpha}(0)=0}^{\phi_{\alpha}(\tilde{t})=0}
        {\cal D}\phi_{\alpha}(\tau) \;
        \exp\Biggl\{
        -\frac{1}{2}\int_{0}^{\tilde{t}} d \tau
   \Biggl[
   \zeta_{*} \, \sum_{\alpha=1}^{K} \Bigl(\partial_\tau \phi_{\alpha}\Bigr)^2
   \; - \;
   \zeta_{*}^{2}
    \sum_{\alpha\not=\beta}^{K} U_{0}\bigl(\phi_{\alpha} - \phi_{\beta}\bigr)
   \Biggr]
        \Biggr\}
\end{equation}
Let us extract from this partition function the explicit contribution containing the linear in time  free energy 
$f_{0}(\zeta_{*})\tilde{t}$ and redefine
\begin{equation}
 \label{40}
{\cal Z}\bigl(\zeta_{*}; K, \tilde{t}\bigr) \; = \; \exp\bigl\{- K \, f_{0}(\zeta_{*})\, \tilde{t} \bigr\} \, \times \,
\tilde{{\cal Z}}\bigl(\zeta_{*}; K \tilde{t}^{1/3}\bigr)
\end{equation}
Here by analogy with the solution of the corresponding
problem with the $\delta$-interactions in the limit $\tilde{t} \to \infty$ the  function 
$\tilde{{\cal Z}}\bigl(\zeta_{*}; K \tilde{t}^{1/3}\bigr)$ is expected to depend on the replica parameter $K$ and the time $\tilde{t}$ 
in the combination $K \tilde{t}^{1/3}$. It is this function which defines the probability distribution function 
$P\bigl({\cal F}\bigr)$   of the fluctuating part of the free energy, eq.(\ref{32}). According to the 
relations, eq.(\ref{5}), (\ref{29}) and (\ref{40}), and the definition (\ref{32}) the  
probability distribution function $P\bigl({\cal F}\bigr)$ and the partition function $\tilde{{\cal Z}}\bigl(K \tilde{t}^{1/3}\bigr)$
are related via the  Laplace transform:
\begin{equation}
 \label{41}
\int_{-\infty}^{+\infty} d{\cal F} \,P\bigl({\cal F}\bigr)\, \exp\bigl\{- K \tilde{t}^{1/3} \, {\cal F}  \bigr\} \; = \; 
\tilde{{\cal Z}}\bigl(\zeta_{*}; K \tilde{t}^{1/3}\bigr)
\end{equation}
which, at least formally, allows to reconstruct the probability distribution function $P\bigl({\cal F}\bigr)$
via the inverse Laplace transform
\begin{equation}
 \label{42}
P\bigl({\cal F}\bigr) \; = \; \int_{-i\infty}^{+i\infty} \frac{ds}{2\pi i} \, \tilde{{\cal Z}}(\zeta_{*}; s) \, 
\exp\bigl\{ s \, {\cal F}  \bigr\}
\end{equation}

\section{Discussion}

Summarizing all the above speculations, the systematic solution of the considered problem in the zero-temperature limit
consists of three steps.

{\bf First}. For a given integer $K$  and for a given real positive $\zeta$ one has to compute the partition
function ${\cal Z}(\zeta; K, t)$, eq.(\ref{39}), 
where $U_{0}\bigl(\phi\bigr) \; = \; \exp\{-\frac{1}{2}\phi^{2}\}$. This partition function in the limit $t \to \infty$
(similar to the case of the $\delta$-interactions)  is expected to factorize into two essentially different contributions:
(a) the one which explicitly reveal the linear in time non-random (self-averaging) free energy part $f_{0}(\zeta)$; and 
(b) a function $\tilde{{\cal Z}}\bigl(\zeta; K t^{1/3} \bigr)$ which depends on the replica parameter $K$ and time $t$
in the combination $K t^{1/3}$, eq.(\ref{40}).

{\bf Second}. The physical value of the parameter $\zeta_{*}$ is defined by the solution of the equation (\ref{37}).
This solution has to be substituted into the function  $\tilde{{\cal Z}}\bigl(\zeta_{*}; K t^{1/3} \bigr)$.

{\bf Third}. The probability distribution function $P\bigl({\cal F}\bigr)$ of the free energy fluctuating part ${\cal F}$
is defined by the Laplace transform relation, eq.(\ref{41}). Here the function $\tilde{{\cal Z}}\bigl(\zeta_{*}; K t^{1/3} \bigr)$
has to be analytically continued for arbitrary complex values of the replica parameter $K$. Then, introducing the Laplace transform 
parameter $s = K t^{1/3}$ the probability distribution function $P\bigl({\cal F}\bigr)$ is obtained by the inverse 
Laplace transform, eq.(\ref{42}). 

As in the considered zero-temperature limit the partition function (\ref{39}) does not depend on the temperature,
the free energy probability distribution function $P\bigl({\cal F}\bigr)$ must be  temperature independent too.
The only "little" problem in the above derivation scheme is that for the moment the solution for  the 
partition function ${\cal Z}(\zeta; K, t)$, eq.(\ref{39}) (the First step) is not known.
Nevertheless even at present (somewhat speculative) stage we can claim that  in the limit $T \to 0$ 
the considered system reveals  the one-step replica symmetry breaking structure, eqs.(\ref{27}) and (\ref{29}),
similar to the one which takes place in the Random Energy Model. Besides,  at the temperature
$T_{*} \sim (u R)^{1/3}$ we observe a crossover from the high- to the low-temperature regime. 
Namely,  at high temperatures, $T \gg T_{*}$, the model is equivalent to the one with the 
$\delta$-correlated potential in which the non-universal prefactor of the fluctuating part of the free 
energy is proportional to $(u/T)^{2/3}$,  
while at $T \ll T_{*}$ this non-universal prefactor  saturates 
at a finite (temperature independent) value $\sim (u^{2}/R)^{1/9}$. 
The formal proof that the zero temperature limit free energy distribution function
of the considered system is indeed the Tracy-Widom one requires further investigation.

\acknowledgments

The main idea of the present research, namely the splitting replicas into the one-step replica symmetry breaking structure
in the zero-temperature limit, belong to Sergei Korshunov. 

I am grateful to Vivien Lecomte, Elisabeth Agoritsas and Herbert Spohn  for numerous illuminating discussions.
An essential part of this work was done during the workshop "New approaches to non-equilibrium
and random systems: KPZ integrability, universality, applications and experiments"
(Jan 11 - Mar 11, 2016)
at Kavli Institute of Theoretical Physics, University of California, Santa Barbara.
This research was supported in part by the National Science Foundation under Grant No. NSF PHY11-25915.

\end{document}